\documentclass[twocolumn]{aastex631}
\usepackage{subfigure}
\usepackage{graphicx}
\usepackage{CJK}
\usepackage{amsmath}
\usepackage{hyperref}
\usepackage{float}
\usepackage{placeins}
\usepackage{needspace}

{\newif\ifnotend
\notendtrue
\def\veclist{ABCDEFGHIJKLMNOPQRSTUVWXYZabcdefghijklmnopqrstuvwxyz.}
\def\top#1#2.{#1}
\def\tail#1#2.{#2.}
\loop\expandafter\xdef\csname v\expandafter\top\veclist\endcsname%
{{\noexpand\bf\expandafter\top\veclist}}
\edef\veclist{\expandafter\tail\veclist}
\if\veclist.\notendfalse\fi\ifnotend\repeat}
{\newif\ifnotend
\notendtrue
\def\callist{ABCDEFGHIJKLMNOPQRSTUVWXYZ.}
\def\top#1#2.{#1}
\def\tail#1#2.{#2.}
\loop\expandafter\xdef\csname c\expandafter\top\callist\endcsname%
{{\noexpand\cal\expandafter\top\callist}}
\edef\callist{\expandafter\tail\callist}
\if\callist.\notendfalse\fi\ifnotend\repeat}

\def\d{{\rm d}}

\def\apjs{ApJS}
\def\apjl{ApJL}

\def\aap{A\&A}

\def\Gyr{\,\mathrm{Gyr}}

\def\kpc{\,\mathrm{kpc}}

\def\kms{\,\mathrm{km\,s}^{-1}}

\def\msun{\,{\rm M}_\odot}

\def\e{\mathrm{e}}

\def\eg{{ e.g.,\ }}
\def\sun{\odot}

\def\Omegab{\Omega_{\rm b}}

\def\p{\partial}\def\i{{\rm i}}

\defcitealias{li2024}{Li24}
\defcitealias{Binney2023}{BV23}
\defcitealias{Binney2024}{BV24}

\begin{document}
\begin{CJK*}{UTF8}{gbsn}
\title{The Rotating Bulge and Halo in the Milky Way: Evidence of Angular Momentum Transferred from the Decelerating Bar}

\author[0000-0002-1126-9289]{Zhuohan Li (李卓翰)}
\altaffiliation{Joint first authors}
\affiliation{National Astronomical Observatories, Chinese Academy of Sciences, Beijing 100101, People's Republic of China}
\affiliation{School of Astronomy and Space Science, University of Chinese Academy of Sciences, Beijing 100049, People's Republic of China}
\author[0000-0001-9949-0625]{Chengdong Li (李承东)}
\altaffiliation{Joint first authors}
\altaffiliation{Corresponding authors. E-mail: \href{mailto:chengdong.li@nju.edu.cn}{chengdong.li@nju.edu.cn}}
\affiliation{School of Astronomy and Space Science, Nanjing University, Nanjing 210093, People's Republic of China }
\affiliation{Key Laboratory of Modern Astronomy and Astrophysics (Nanjing University), Ministry of Education, Nanjing 210093, People's Republic of China }
\author[0000-0002-8980-945X]{Gang Zhao (赵刚)}
\altaffiliation{Corresponding authors. E-mail: \href{mailto:gzhao@nao.cas.cn}{gzhao@nao.cas.cn}}
\affiliation{National Astronomical Observatories, Chinese Academy of Sciences, Beijing 100101, People's Republic of China}
\affiliation{School of Astronomy and Space Science, University of Chinese Academy of Sciences, Beijing 100049, People's Republic of China}
\author[0009-0008-1319-1084]{Ruizhi Zhang (张睿之)}
\affiliation{National Astronomical Observatories, Chinese Academy of Sciences, Beijing 100101, People's Republic of China}
\affiliation{School of Astronomy and Space Science, University of Chinese Academy of Sciences, Beijing 100049, People's Republic of China}
\author[0000-0002-0642-5689]{Xiang-Xiang Xue (薛香香)}
\affiliation{National Astronomical Observatories, Chinese Academy of Sciences, Beijing 100101, People's Republic of China}
\affiliation{Institute for Frontiers in Astronomy and Astrophysics, Beijing Normal University, Beijing 102206, Peopleʼs Republic of China}



\begin{abstract}
Recent observations indicate that both the Milky Way bulge and inner halo exhibit angular momentum, although the origin and evolution of this prograde signature remain ambiguous. One plausible scenario involves secular evolution induced by the central bar and spiral arms. In this study, we identified a component consisting of 1,175,737 stars with net rotation through the application of a neural network (NN) method. To investigate the composition of this rotating sample and the origin of its rotation, we conducted a test particle simulation incorporating an equilibrium axisymmetric background potential together with a central decelerating bar. The test particles were generated using a distribution function (DF) model derived from observational constraints. Our results indicate that the decelerating bar transfers angular momentum to the pseudo-stars, and the rotational profile from our simulation shows strong agreement with observational data. These findings suggest that the rotating sample identified by our NN model predominantly comprises bulge, halo, and thick disk stars, and that the central decelerating bar is pivotal in shaping the inner Galaxy's kinematics through angular momentum transfer.
\end{abstract}



\section{Introduction} \label{sec:intro}

In disk galaxy formation scenarios, the bulge is traditionally assumed to have no net rotation or angular momentum (see Chapter 11.2.4 in \citealt{Mo2010}). However, recent observational studies have revealed that the Milky Way (MW) bulge exhibits measurable net rotation, even among its metal-poor stellar populations \citep[see \eg][]{Binney2024,Arentsen2024}.  
Similarly, the stellar halo is believed to possess non-zero angular momentum, particularly in the inner regions, where a net rotation of approximately $v_{\phi}=30\sim40\kms$ has been reported \citep{Wegg2019,Li2022a}. The outer halo of the MW is dominated by the merger and accretion remnants which may inherit the angular momentum from their progenitors \citep{Ibata1994,Belokurov2018,Myeong2019}. Nevertheless, the mechanism responsible for transferring angular momentum to the bulge and inner halo remains unclear. Secular evolution, which is triggered by the global instabilities, such as the bar and spiral arms, is usually considered as one of the key processes in facilitating angular momentum transfer \citep{Sellwood2002,Sellwood2014}.

Dynamical friction has long been recognized as one of the primary mechanisms facilitating angular momentum exchange between the central bar and the dark matter halo \citep[see \eg][]{Chandrasekhar1943,Tremaine1984,Chiba2023,Hamilton2023}. Moreover, it serves as a significant piece of evidence for the existence of dark matter, as its effects are inherently dependent on the properties of the dark matter \citep{Debattista2000,Sellwood2016}. There are also observational evidence in the Milky Way that dynamical friction plays an important role in redistributing angular momentum between the decelerating bar and the baryons \citep{Chiba2021a,Dillamore2023,Li2023b,Yuan2023,zhang2025observational}, as well as in driving the orbital decay of satellite galaxies \citep{Lin1983,Weinberg1986}. In realistic stellar systems, where periodic interactions occur between perturbers and individual particles, the most effective mechanism for angular momentum transfer is through orbital resonances \citep{Tremaine1984, Chiba2023}, which has been numerically confirmed in the formation scenario of the very metal-poor thin disk stars in the solar vicinity due to the bar's corotation resonance \citep{Li2023b}.\par

The third data release of $Gaia$ \citep{GaiaDR3} provided astrometric data for over 1.8 billion stars, with more than 30 million stars having 6D phase space information, presenting an ideal opportunity to further study the aforementioned issues in detail through observation. To manage this vast amount of data, deep learning methods can be employed. \citet[hereafter \citetalias{li2024}]{li2024} developed an neural network (NN) methodology to distinguish between ex-situ and in-situ stellar populations within the $Gaia$ DR3 catalogue. Their NN was applied to a target sample of 27 million stars, all of which had relatively precise radial velocity from $Gaia$ DR3 and photo-astrometric distance provided by \citet{anders2022photo}. This sample was subsequently divided into two distinct categories, corresponding to the in-situ and ex-situ stars respectively. While the primary focus of their investigation was the ex-situ components, mainly encompassing dwarf galaxies and substructures, the in-situ sample also possesses significant research potential. Notably, the NN model does not directly classify stars, but maps the input parameters to a value, the distribution of which contains the kinematic information of the stars. In this way, we can use the output values of the NN to study the kinematic patterns of in-situ stars. Furthermore, in-situ stars directly reflect the characteristics of the MW itself and are not affected by the contamination of accretion remnant. In light of this, our research endeavors to shift the focus towards the in-situ part of the samples characterized in \citetalias{li2024}.\par

In this study, we identified a subset of stars exhibiting atypical rotational characteristics within $Gaia$ DR3, providing evidence of net rotation in both the bulge and halo. Our results suggest that a decelerating bar, capable of transferring angular momentum to bulge and halo stars, may be responsible for this phenomenon. The modeling and simulations were performed using \texttt{AGAMA} \citep{Vasiliev2019}, employing the distribution function (DF) modeling approach. The DF can be taken as a function of isolating integrals in a given gravitational field \citep{Jeans1915}, which is most conveniently parameterized in terms of three actions: $J_R$, representing the amplitude of radial excursions; $J_z$, quantifying vertical oscillations relative to the Galactic plane; and $J_\phi$, corresponding to the angular momentum component aligned with the Milky Way’s assumed symmetry axis. The model we adopted consists of a dark matter halo, a stellar halo, a bulge and four disk components: a young disk, a middle-aged disk, an old thin disk, and a high-$\alpha$ disk for the MW \citep{Li2022a,Binney2023}. The gravitational potential generated by the DFs can thus be computed iteratively using the methods introduced in \citet{Binney2014}. The initial pseudo-stars in our simulation were generated by the DF modelling method, and their orbital integration was then computed through an axisymmetric background potential plus a decelerating bar via \texttt{AGAMA}. \par

The paper is organized as follows. In Section~\ref{sec:data}, we present the observational results, including the identification of an atypical rotating stellar component. Section~\ref{sec:model} describes the DF model used to generate the initial pseudo-star population and construct the total gravitational potential of the Milky Way, along with the bar model implemented in the test-particle simulation. Our simulation results and their comparison with observational data are discussed in Section~\ref{sec:results}. Finally, Section~\ref{sec:conclusions} provides a summary of our findings and outlines potential directions for future research.

\section{Observational sample} \label{sec:data}
 Astrometric data from $Gaia$ \citep{TheGaiamission} facilitate the kinematic study of the MW. The inclusion of radial velocity information in the latest data release further enhances the potential for detailed dynamical investigations. Following the workflow in \citetalias{li2024}, we curate a sample from $Gaia$ DR3 for our analysis. Specifically, we selected stars with available photo-astrometric distance and radial velocity within $Gaia$ DR3 \citep{anders2022photo, GaiaDR3}, imposing specific criteria to ensure data quality. These criteria include a renormalized unit weight error (RUWE) of less than 1.4 \citep{lindegren2021gaia}, a radial velocity error of less than 20 km s$^{-1}$, and a distance error of less than 30\%. \par
As mentioned in \citetalias{li2024}, the distances of a limited number of stars located behind the Galactic center and in proximity to the Galactic plane may be overestimated, potentially due to extinction issues. Consequently, we exclude stars within the region defined by the following parameters: $x$ $>$ 0 kpc, $z$ $>$ $-$3 kpc, $r\rm_{gc}$ $>$ 6 kpc, $-$50$^{\circ}$ $<$ $l$ $<$ 50$^{\circ}$ and $-$10$^{\circ}$ $<$ $b$ $<$ 15$^{\circ}$. Here, $r\rm_{gc}$ denotes the Galactocentric distance in the spherical coordinate system, while $l$ and $b$ represent the Galactic longitude and Galactic latitude, respectively. Furthermore, we eliminate stars with positive total energy, resulting in a target sample of 27,085,748 stars.
\subsection{Neural network methodology} \label{sec:NN method}
Adhering to the same architectural design as delineated in \citetalias{li2024}, we construct an NN model. The NN architecture includes two distinct input heads. The first input head processes three-dimensional position and velocity data in the Galactocentric Cartesian coordinate system, while the second input head is designed to process the action variables $J_{R}$, $J_{z}$, and $J_{\phi}$.\par
We adopt a similar training framework as \citetalias{li2024}, which comprises an initial training phase predicated on the FIRE-2 simulation \citep{hopkins2018fire}, followed by a fine-tuning process based on observational data. During the first training phase, we employ the latest released synthetic $Gaia$ DR3 survey catalogue based on FIRE-2 simulation \citep{nguyen24}, rather than the former version \citep{sanderson2020synthetic} utilized by \citetalias{li2024}. Our dataset consists of the first five slices of the lsr-0 dataset from the m12i catalogue, spanning a heliocentric distance range of 8 kpc. In the second training phase, we utilize the same training data as \citetalias{li2024}.\par

The NN model establishes a mapping from the input kinematic parameters to a continuous predicted value, which spans the interval from 0 to 1 and can be interpreted as a probability. A prediction approaching 1 suggests that the NN model exhibits a stronger propensity to categorize a star as ex-situ. Conversely, a predicted value that approaches 0 indicates an in-situ origin for the star. Furthermore, the prediction contains the input information of the stars, hence the predicted values for stars with similar kinematic patterns will also be close.

\subsection{The detection of an atypical rotating component}
\begin{figure}
 \includegraphics[width=\columnwidth]{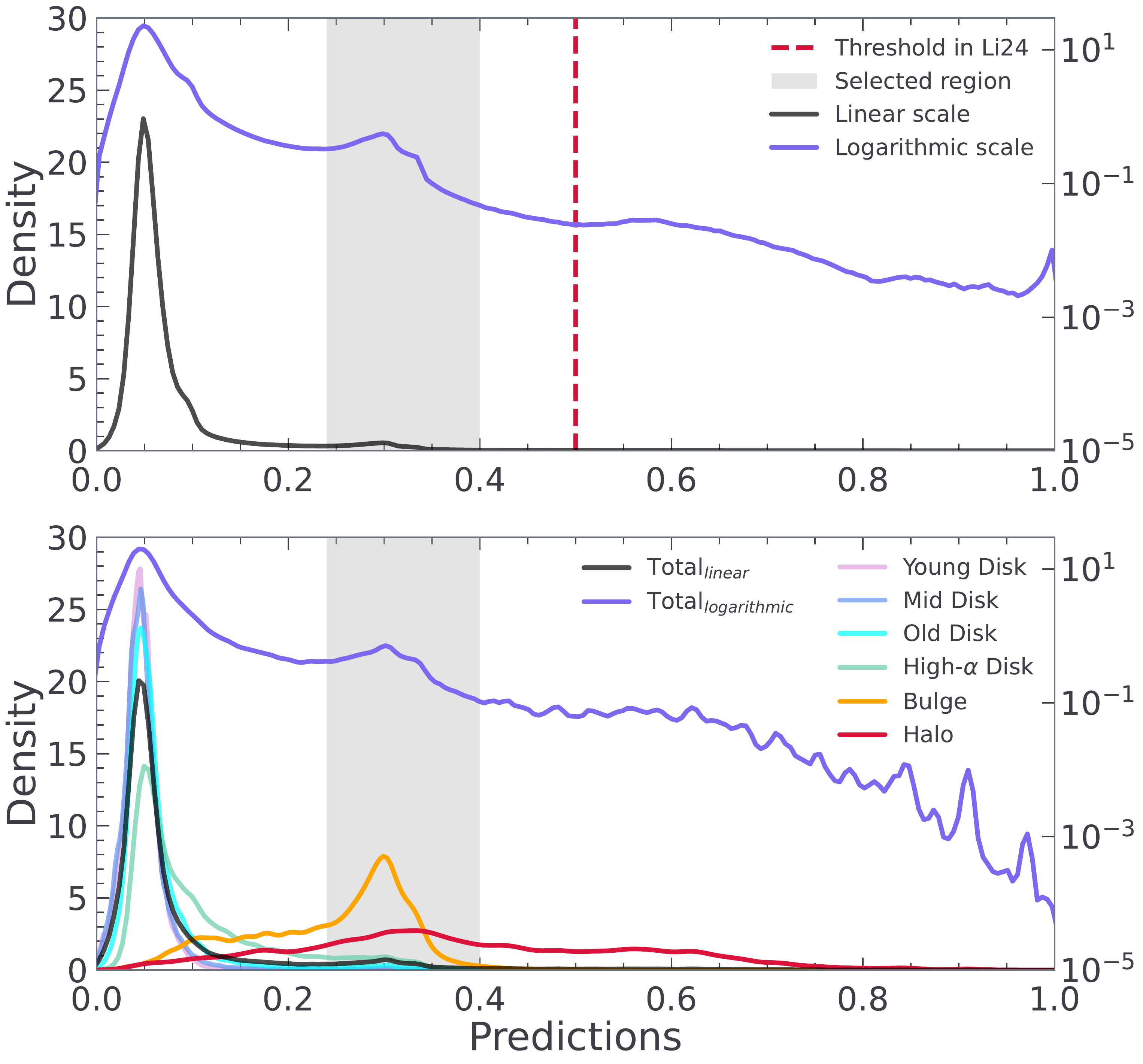}
 \caption{Upper panel: neural network predictions for the $Gaia$ DR3 sample. We show the kernel density distribution on both linear and logarithmic scales. The dashed line marks the threshold in \citet{li2024}. The area of interest for this study, specifically the range with prediction values between 0.24 and 0.4, is highlighted with a shaded background. Lower panel: neural network predictions for the resampled simulation data. The kernel density distribution of the total sample is shown on both linear and logarithmic scales, while the distributions of individual components are plotted on a linear scale using distinct colors.}
 \label{fig:predictions}
\end{figure}

Upon completion of the training phase, we apply the NN model to the $Gaia$ DR3 sample. The distribution of the NN predictions is depicted in Figure~\ref{fig:predictions}, where \citetalias{li2024} established a threshold of 0.5, classifying stars with predictions exceeding this value as ex-situ. Using a single threshold to divide the sample is feasible for identifying ex-situ stars, as they are relatively rare and far from the main peak of the sample distribution. However, for in-situ stars, a more detailed distinction is required.\par
Within the in-situ segment, a prominent peak is evident, corresponding to the thin and thick disks. Additionally, a secondary peak can be observed, whose composition remains ambiguous. To extract this component, we delineate a slice between the predictions of 0.24 and 0.4 which contains 1,175,737 stars, as highlighted by the background in Figure~\ref{fig:predictions}. The left boundary is defined by the valley bottom to the left of the secondary peak, while the right boundary is selected to exclude the member stars of GSE \citep{Belokurov2018, helmi2018merger, myeong2018sausage, haywood2018disguise}. \par

The selected component has no obvious kinematic tendency in the vertical and radial directions, since the median $V_{R}$ and the median $V_{z}$ are close to zero, as depicted in Figure~\ref{fig:velocity_obs}. Nevertheless, it exhibits a net rotation of approximately 80 km s$^{-1}$. This rotation speed is substantial but not comparable to either the thin disk or the thick disk. In terms of spatial distribution, as shown in the upper panels of Figure~\ref{fig:spatial}, about 90\% of stars are located in the area where the Galactocentric distance in the cylindrical coordinate system ($R$) is less than 5 kpc, and the absolute value of the vertical distance from the Galactic disk ($z$) for over 80\% of stars exceeds 600 pc. Consequently, we do not consider this in-situ rotating component to be the disk, but a composite of the bulge and halo. To verify this conjecture, we subsequently carry out test-particle simulations.

\section{Modelling frameworks and simulation scheme}{\label{sec:model}}  
To investigate the origin of this atypically rotating component, we conduct a simulation using test particles initialized with a DF model and evolved within a predefined non-axisymmetric potential, which includes a central decelerating bar to emulate the evolution of the MW. This approach allows for flexible adjustments to the perturbation models and enables efficient simulations with eight million test particles within a few hours. Moreover, at any point during the simulation, the analytical models and the test particle responses to the perturbations are fully accessible. In this section, we present the DF model used to generate pseudo-stars and describe the non-axisymmetric potential in detail.

\subsection{The bar model}{\label{sec:bar}}
We adopt a decelerating bar model similar to that used in \citet{Li2023}, which is based on the Made-to-Measure model by \citet{Portail2017}, with updated parameters from \citet{Sormani2022}. The key difference from previous studies is that we constrain the decelerating bar to terminate at a phase angle of $\phi\,=\,28^{\circ}$ relative to the Sun, aligning with observations of the present-day morphology of the MW \citep{Wegg2015}. In our model, the pattern speed of the bar decreases from an initial value of $\Omegab\,=\,-56\,\kms\,\kpc^{-1}$ to $\Omegab\,=\,-35\,\kms\,\kpc^{-1}$ at the present time (e.g. \citealt{Chiba2021b, Clarke2022, Zhang2024kinematics}). 

\begin{figure}
 \includegraphics[width=\columnwidth]{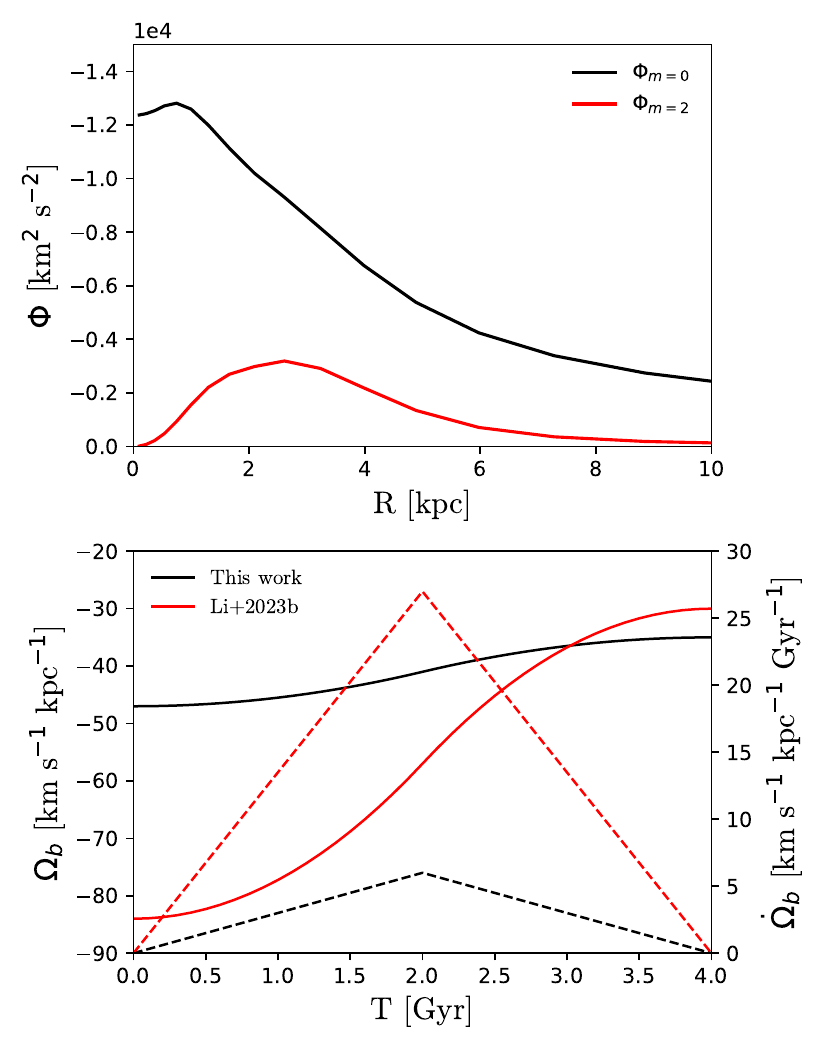}
 \caption{Upper panel: the m=0 and m=2 Fourier terms as a function of distance at z = 0 for the slowing down bar model in black and red respectively. Lower panel: the pattern speed $\Omegab$ in this work is shown in solid black curve, and the decelerating rate $\dot \Omega_{\rm b}$ in this work is shown in dashed black curve. In order to make a comparison, the  pattern speed $\Omegab$ and the decelerating rate $\dot \Omega_{\rm b}$ in \citet{Li2023b} are shown in red solid and dashed lines respectively.}
 \label{fig:fourier}
\end{figure}

The upper panel of Figure~\ref{fig:fourier} shows the $m=0$ and $m=2$ Fourier components as a function of radius at $z = 0$ for the slowing-down bar model, plotted in black and red, respectively. The lower panel compares the pattern speed and its deceleration rate over time between this work and \citet{Li2023b}. The pattern speed $\Omegab$ in this work is plotted as a solid black curve, and its deceleration rate $\dot \Omega_{\rm b}$ as a dashed black curve. The corresponding quantities from \citet{Li2023b} are shown as red solid and dashed curves, respectively. In our simulation, the pattern speed decreases by approximately $37.5 \%$, which is consistent with the predictions of \citet{Chiba2021b} and falls within the slow-bar regime. In this regime, a significant fraction of stars are expected to be trapped in the bar resonances (see also the discussions in \citealt{Tremaine1984, Chiba2023, Li2023b}). Throughout the simulation, the mass and radial extent of the bar evolve continuously, reaching 2.0 and 1.26 times their initial values, to approximately mimic the growth of the bar.

It is important to note that, throughout this study, we neglect the self-gravity of the disk. As demonstrated by \citet{Weinberg1989}, the resonant structure can be strengthened via self-gravitating processes \citep[see also the discussions in][]{Dootson2022,Chiba2023}. Consequently, self-gravity is expected to influence the diffusion of the disk. However, incorporating such effects is beyond the scope of this work, which focuses on a preliminary qualitative analysis rather than a detailed quantitative investigation.

The orbital integration routine of \texttt{AGAMA} is adopted to compute the trajectories of all the mock stars over a total duration of 4 $\Gyr$, with trajectory data recorded at intervals of 0.02 $\Gyr$. The bar is added to the total potential at $T = 0.0 \Gyr$ with initial radial profile and a pattern speed of $\Omegab\,=\,-56\,\kms\,\kpc^{-1}$. Then the bar starts to decelerate and grow in mass and radial direction, reaching a pattern speed of $\Omegab\,=\,-35\,\kms\,\kpc^{-1}$ at the end of the simulation.

\subsection{Synthetic observation}{\label{sec:selection}}
In an effort to better simulate the $Gaia$ data, observational uncertainties are systematically integrated into the simulation framework. Initially, we transform the Cartesian coordinates in the simulation into the celestial coordinate system, assuming the circular speed at the Sun as 232.8 km s$^{-1}$ and the Solar Galactocentric distance as 8.2 kpc \citep{mcmillan2016mass}. The distance from the Sun to the Galactic plane is set at 20.8 pc \citep{Bennett19}. For the solar peculiar motion, we adopt the value reported by \citet{schonrich2010local}, yielding (U, V, W) $=$ (11.1, 12.24, 7.25) $\kms$. Within our $Gaia$ DR3 sample, the typical uncertainties for right ascension (ra), declination (dec), proper motion in right ascension (pmra), and proper motion in declination (pmdec) are 0.013 mas, 0.012 mas, 0.016 mas yr$^{-1}$, and 0.015 mas yr$^{-1}$, respectively. For radial velocity, we apply a typical error of 2 km s$^{-1}$. The distance measurements are assigned a typical relative error of 10\%. Corresponding to the number of particles in the simulation, deviations for the first five observational parameters are sampled from Gaussian distributions with zero mean and standard deviations equivalent to the typical errors, which are then added to the aggregate simulation dataset. Distance values are derived similarly, except for being sampled from log-normal distributions.\par
Furthermore, we implement a three-dimensional resampling on the simulation data to model the selection effect in the $Gaia$ sample. Our analysis concentrates on the specified region where $-$20 kpc $\leq x \leq$ 10 kpc, $-$15 kpc $\leq y \leq$ 15 kpc, $-$10 kpc $\leq z \leq$ 10 kpc. Within this space, the $Gaia$ DR3 sample is partitioned into cubic segments with a step size of 0.5 kpc for the x and y dimensions, and 0.2 kpc for the $z$ dimension. For each cubic segment, the number of observed stars is normalized by the total stellar count across the entire volume, yielding a probability density function (PDF) for the spatial distribution of stars in the $Gaia$ sample. An identical grid is then applied to the simulation dataset, where the same PDF is used to weight the simulated stellar counts, thereby estimating the expected number of stars per cubic segment. Subsequently, a stochastic sampling is conducted within each segment to approximate the observed distribution. After the sampling, the fractions of the halo, bulge, high-$\alpha$ disk, old disk, middle disk, and young disk become 5.2\%, 3.1\%, 19.5\%, 19.5\%, 34.9\%, and 17.8\%, respectively.\par

\section{Results and discussions}\label{sec:results}
\subsection{Neural network selection in the simulation}
The resampling process ensures that the spatial distribution of the simulated samples closely mirrors that of the observed samples. This alignment allows for the direct application of the neural network, which was trained and applied on the observational data, to the simulation data. The error-convolved three-dimensional coordinates and velocities are adopted as inputs for the neural network. The input actions are recalculated based on the coordinates and velocities under the potential provided by \citet{mcmillan2016mass}, which was utilized during the training process. Subsequent analyses will also be based on these recalculated actions. \par

The distribution of NN prediction values for the simulated sample is presented in the lower panel of Figure~\ref{fig:predictions}. Overall, the simulation and observation are in good agreement, with the black and blue curves showing close correspondence. Notably, the disk components exhibit a single peak around 0.05, which aligns with the prominent peak observed in the upper panel. The bulge displays two peaks: the primary peak lies within the prediction value range of 0.24 to 0.4, while the secondary peak is around 0.1, indicating stronger rotation. The distribution of the halo is relatively uniform, with the main peak also falling between the predicted values of 0.24 and 0.4. Within the predicted value range of 0.24 to 0.4, halo stars constitute 31.0\%, bulge stars account for 28.4\%, and high-$\alpha$ disk stars represent 27.2\%. Given the context of our simulation, we infer that in the observational data, the bulge and halo may also contribute to the second peak observed in the upper panel.\par

\begin{figure*}
 \includegraphics[width=\textwidth]{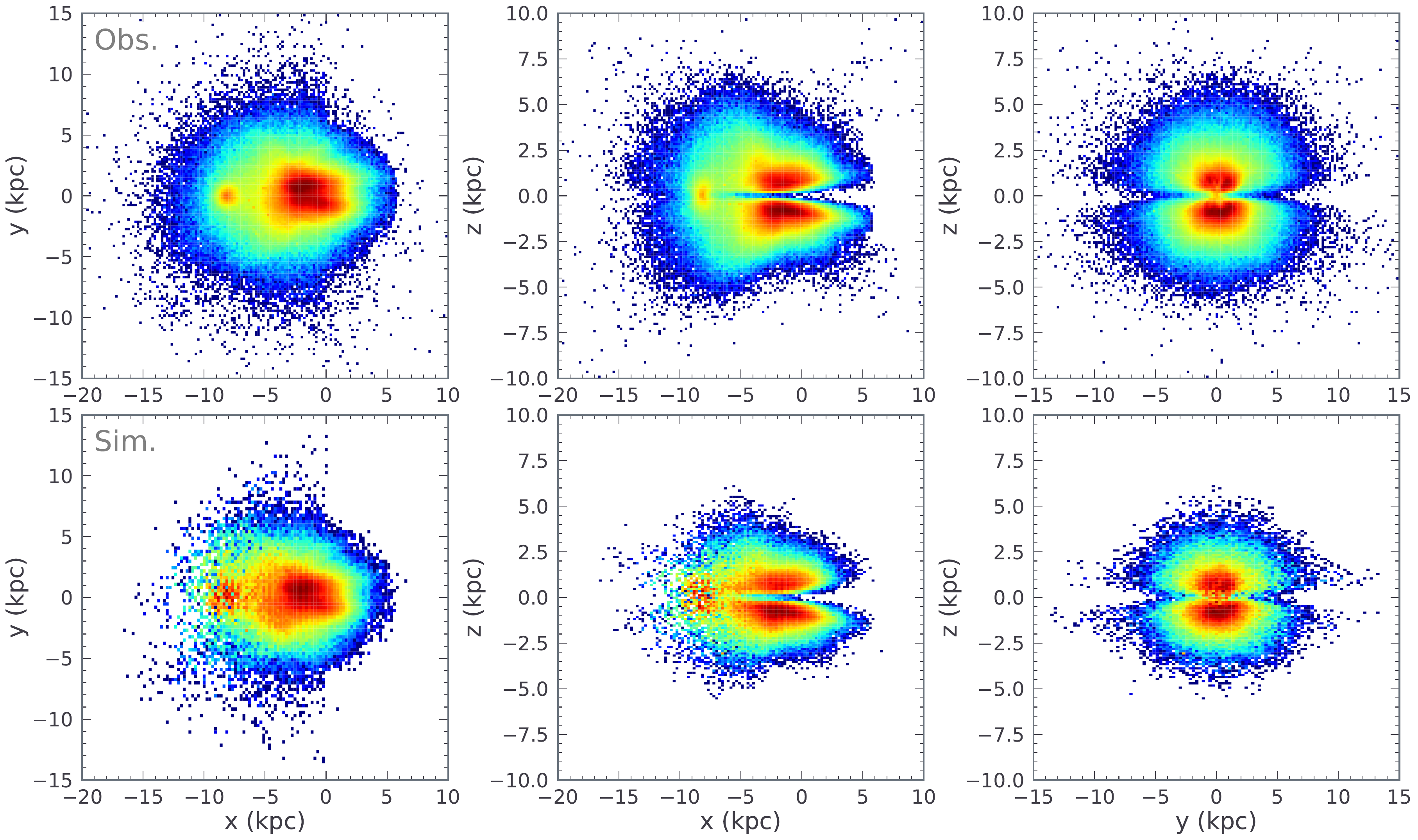}
 \centering
 \caption{Distribution in the coordinate space for stars with NN predictions in the selected range. The upper row represents observational data, while the lower row depicts simulation results. Each subplot is color-coded to indicate the number density of stars, with warmer colors signifying higher densities.}
 \label{fig:spatial}
\end{figure*}

We further examine the spatial distribution of the stars whose predictions are between 0.24 and 0.4 both in the simulation and observation, as illustrated in Figure~\ref{fig:spatial}. On all the three projection planes, the distribution of simulated samples closely resembles that of the observed samples, indicating that the spatial distribution of stars selected by the NN in both simulation and observation is consistent. The overdensity in the solar vicinity observed in the observational samples are attributed to the selection effect of $Gaia$, which is replicated in the simulated samples. This consistency in space between the simulated and observed samples suggests that they likely have similar compositions. Furthermore, the region where the observational samples are located correspond to the area where the bulge and halo are situated. This result further supports our hypothesis that bulge and halo stars contribute to the part within the prediction value range of 0.24 to 0.4 as determined by the NN in the observational data.\par

\subsection{The rotating bulge and halo}\label{sec:ang}
\begin{figure}
 \includegraphics[width=\columnwidth]{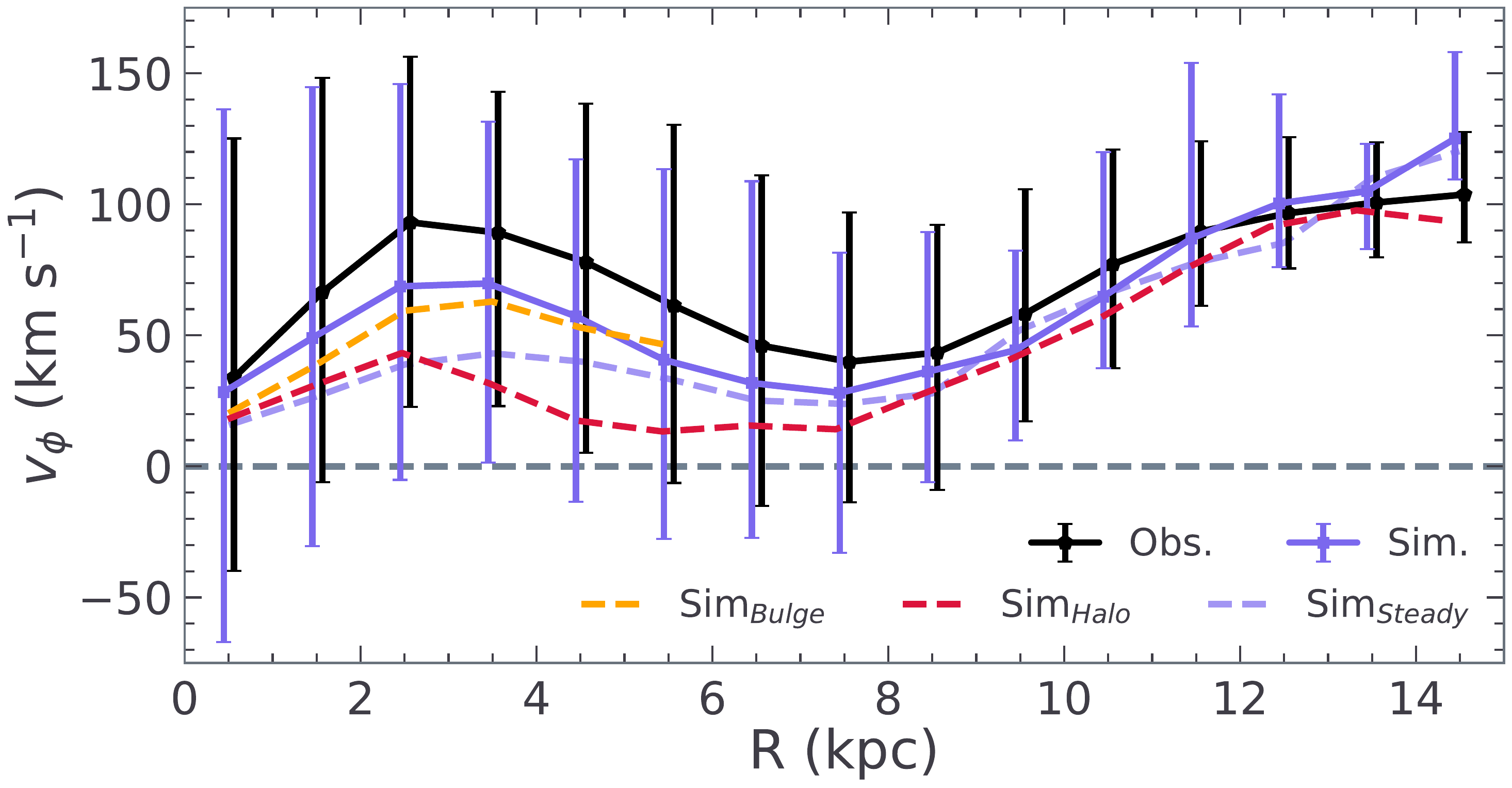}
 \caption{The variation of $v_{\phi}$ with $R$ for simulated and observed samples. The dots represent the median $V_{\phi}$ values in each bin, with error bars indicating the 16\textsuperscript{th} and 84\textsuperscript{th} percentiles. The solid lines illustrate a comparison of all simulated samples with NN prediction values ranging between 0.24 and 0.4, against the observed samples within the same range. The orange and red dashed lines depict the distributions of bulge and halo components within the simulated sample, respectively. For comparison, a simulation result incorporating a steadily rotating bar is shown as a blue dashed line. Small offsets in $x$-axis are applied for visual clarity.}
 \label{fig:rvphi}
\end{figure}

We analyze the variation of $v_{\phi}$ with $R$ for both simulated and observed samples, as illustrated in Figure~\ref{fig:rvphi}. The solid black curve represents the observed sample with NN prediction values ranging from 0.24 to 0.4, while the solid blue curve corresponds to all simulated samples within the same NN prediction range. Our simulation results exhibit strong agreement with the observed data in Figure~\ref{fig:rvphi}. The curve, in both observations and simulations, exhibits a distinct trend: it increases from the Galactic center to approximately 3 kpc, where the influence of the Galactic bar could be most pronounced. Beyond this point, the curve declines until reaching the solar radius, where the disk potential is expected to dominate \citepalias{Binney2023} and the bulge is already truncated \citepalias{Binney2024}. However, in the selected sample, our NN excludes the majority of the disk stars, thereby isolating a predominantly halo component. Moreover, the flattening parameter of the stellar halo, $q$, attains its lowest value around the solar radius \citep{Li2022a}, which enhances the observational detectability of halo stars at this location. Consequently, halo stars dominate the sample in the solar vicinity. Beyond the solar radius, the curve continues to increase, mainly reflecting the influence of halo stars in this region.

The orange and red dashed lines depict the velocity variation of simulated bulge and halo components with $R$, respectively. The orange curve, indicating a slightly rotating bulge, ends at $R \sim$ 6$\kpc$, which is in accordance with idea of the truncated bulge model in \citetalias{Binney2024}. The increase of the bulge curve from the Galactic center to $R\sim3\kpc$ is mainly due to the angular momentum transfer from the bar to the bulge particles. The decrease of the curve from $R\sim3\kpc$ to the solar radius is also mainly determined by the bulge, since most of the bulge particles that are trapped by the corotation of the bar and gain substantial angular momentum have NN prediction values below 0.2, they are excluded by the selection criteria and do not contribute to our final result. 

On the other hand, the red curve shows the velocity variation for the halo particles, covering a larger range of radii. The inner part of halo, within $\sim 8\kpc$, exhibits only a slight rotation of approximately $40 \kms$. The rotational motion in our sample is largely driven by halo particles beyond the solar radius. The strong agreement between observations and simulations at large radii suggests that halo stars predominantly contribute to the observational sample beyond the solar radius.

To examine the specific role of bar deceleration, we performed a comparison simulation based on a steadily rotating bar model, following the setup of \citet{Chiba2022}. In this model, the bar maintains a constant pattern speed of $\Omegab = -35 \kms \kpc^{-1}$. As depicted by the dashed blue line, the simulation result exhibits a significant discrepancy from the observed data, highlighting the necessity of including bar deceleration in the model.

\begin{figure*}
 \includegraphics[width=\textwidth]{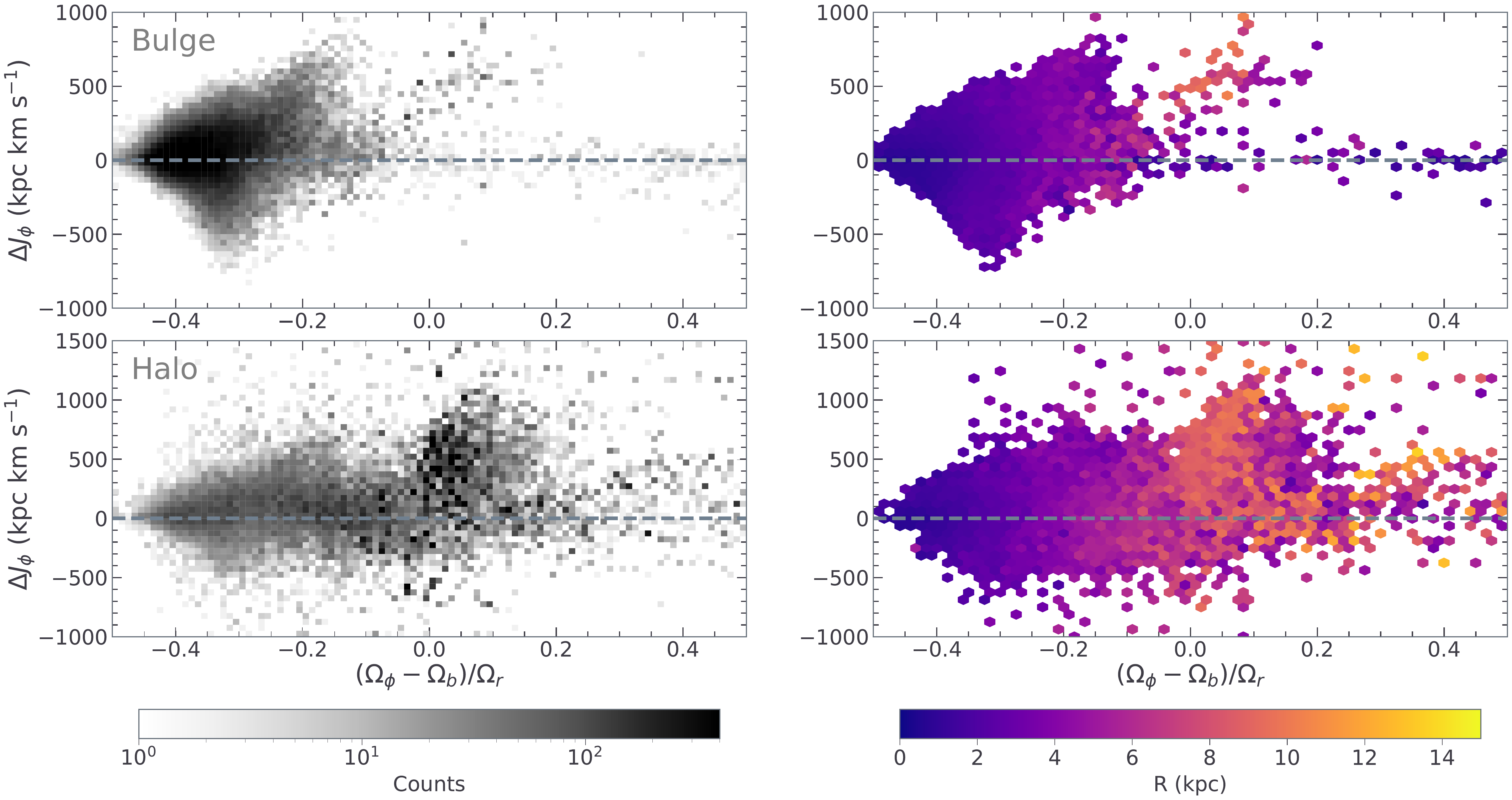}
 \centering
 \caption{Depiction of the angular momentum transfer for bulge stars and halo stars with prediction values between 0.24 and 0.4. The left column is color-coded by the number density, while the right column is color-coded to indicate the Galactocentric distance. The parameters depicted in this figure are not error-convolved, thereby presenting an accurate representation of the conditions within the simulation.}
 \label{fig:jphi_omega}
\end{figure*}

The relation between the change of the angular momentum for test particles between $T=2\Gyr$ to $T=4\Gyr$ and $\epsilon=(\Omega_{\phi}-\Omegab)/\Omega_{r}$ is shown in Figure~\ref{fig:jphi_omega}, where $\Omega_{\phi}$ and $\Omega_{r}$ are the azimuthal and radial frequencies of the test particles and $\Omegab$ is the bar's pattern speed at $T=4\Gyr$. The left column is color-coded by the number density, while the right column is color-coded to indicate the Galactocentric distance. 

The first row presents bulge particles with NN prediction values between 0.24 and 0.4, demonstrating a net gain in angular momentum over the time interval. Notably, these particles are not predominantly trapped by either the corotation resonance $(\epsilon = 0)$ or the Inner/Outer Lindblad Resonances (I/OLR) $(\epsilon = \pm\frac{1}{2})$. This does not imply that bar resonances fail to contribute to angular momentum transfer. On the contrary, bulge particles trapped by the bar's corotation resonance gain substantial angular momentum through secular evolution, leading to prediction values around 0.1 and exhibiting more disk-like orbits.Furthermore, the Galactocentric distances of the bulge particles in this figure are largely confined within $\sim 8\kpc$, with only a minor fraction exceeding this limit near $(\epsilon = 0)$. These outliers are evidently influenced by the bar's corotation resonance. While the corotation resonance is a key driver of radial migration in the Galactic disk (e.g. \citealt{Sellwood2002,Minchev2012,Chiba2021a,Li2023b}), the limited Galactocentric radii of the affected particles in our simulation suggest that significant radial migration does not occur; rather, these particles remain within the bulge region. Nonetheless, they exhibit net rotational gain during the simulation, indicating that beyond major resonances, additional mechanisms could facilitate angular momentum transfer from the bar to baryonic components during dynamical friction and bar deceleration.

The dynamics of halo particles are more complex, as their substantial vertical excursions prevent them from being tightly confined within the Galactic plane. As a result, within several $\kpc$ of the Galactic center, halo particles do not gain angular momentum as efficiently as bulge particles, leading to weaker net rotation, as indicated in Figure~\ref{fig:rvphi}. However, at larger Galactic radii, the situation differs significantly. A substantial fraction of halo particles cluster around $\epsilon = 0$ in the lower-left panel of Figure~\ref{fig:jphi_omega}, signifying resonance trapping by the bar's corotation resonance. These particles experience the largest angular momentum gains and exhibit the most extended Galactocentric radii among the halo population, providing compelling evidence that resonance trapping is the most efficient mechanism for angular momentum transfer. Meanwhile, this halo population contributes to the increase in $v_\phi$ beyond the solar radius, as seen in Figure~\ref{fig:rvphi}.

Overall, bar deceleration due to dynamical friction plays a crucial role in driving the net rotation of both the bulge and halo. Resonance trapping emerges as the dominant pathway for angular momentum transfer. However, in our simulation, particles with NN predictions between 0.24 and 0.4 exhibit only modest rotation compared to disk and bulge particles that gain greater angular momentum. This suggests that minor resonances (other than corotation and ILR/OLR) or alternative processes contribute to the angular momentum evolution of these particles. 

\section{Conclusions}\label{sec:conclusions}
Upon applying a well-trained NN \citep{li2024} to the extensive dataset from $Gaia$ DR3 \citep{GaiaDR3}, we identified a subset of stars exhibiting atypical rotational characteristics. For the first time, we confirmed the net rotation of both the bulge and inner halo on a million-level sample. Our findings indicate that both the bulge and halo exhibit azimuthal velocities ranging from 30 to 40 $\kms$ within the innermost 1 $\kpc$, which increase to over 50 $\kms$ at larger radii.\par
We consider that the angular momentum acquired by the bulge and halo can be provided by the bar, through dynamical friction. The most efficient mechanism for angular momentum transfer is resonance trapping, which imparts angular momentum to halo stars with  $R$ $>$ 8 $\kpc$ and to bulge members with higher azimuthal velocities. Our bulge sample with predicted values between 0.24 and 0.4, appears to have gained angular momentum predominantly through mechanisms other than resonance trapping. Nevertheless, it is evident that the deceleration of the bar due to dynamical friction is the key factor in the angular momentum transfer between the bar and the baryons.\par
According to our simulation, the pattern speed of the bar has decelerated by $\sim$ $37.5\%$ over the last 4 $\Gyr$, and its current mass is approximately two times its initial value. Given the consistency between our simulation results and observational data, we provide new evidence that the bar can transfer angular momentum to the bulge and halo, and reveal that this process is mainly driven by the decelerating of bar.\par
In summary, we present observational evidence indicating that both the bulge and halo exhibit net rotation. We propose a possible interpretation of this phenomenon: the decelerating bar transfers angular momentum to the bulge and halo stars. However, there are caveats in our modeling work. Firstly, our test particle simulations neglect self-gravity, which amplifies the perturbation in the disk \citep{Widrow2023}. Since our focus is on the interaction between the bar and other galactic components in angular momentum transfer, a full treatment of the disk's response to the perturbation due to the self-gravity is left for future works. Secondly, the bar is not the sole feature capable of causing angular momentum transfer. The spiral arms and giant molecular clouds (GMCs) can also facilitate angular momentum transfer, particularly within the disk components. Future research efforts are necessary to elucidate the mechanisms of angular momentum exchange, encompassing not only the interactions between the bar, dark matter, and baryons, but also those involving spiral arms and GMCs, while fully accounting for self-gravity.

\section*{Acknowledgements}
This study is supported by the National Natural Science Foundation of China under grant Nos. 12588202 and 12373020, National Key Research and Development Program of China Nos. 2024YFA1611902 and 2023YFE0107800, CAS Project for Young Scientists in Basic Research grant Nos. YSBR-062 and YSBR-092, and the science research grants from the China Manned Space Project with NO. CMS-CSST-2025-A11.
\par
We have made use of data from the European Space Agency (ESA) mission
{\it Gaia} (\url{https://www.cosmos.esa.int/gaia}), processed by the {\it Gaia}
Data Processing and Analysis Consortium (DPAC,
\url{https://www.cosmos.esa.int/web/gaia/dpac/consortium}). Funding for the DPAC
has been provided by national institutions, in particular the institutions
participating in the {\it Gaia} Multilateral Agreement.
\par
Guoshoujing Telescope (the Large Sky Area Multi-Object Fiber Spectroscopic Telescope LAMOST) is a National Major Scientific Project built by the Chinese Academy of Sciences. Funding for the project has been provided by the National Development and Reform Commission. LAMOST is operated and managed by the National Astronomical Observatories, Chinese Academy of Sciences.
\par
Funding for the Sloan Digital Sky 
Survey IV has been provided by the 
Alfred P. Sloan Foundation, the U.S. 
Department of Energy Office of 
Science, and the Participating 
Institutions.
SDSS-IV acknowledges support and 
resources from the Center for High 
Performance Computing  at the 
University of Utah. The SDSS 
website is \url{www.sdss.org}.

SDSS-IV is managed by the 
Astrophysical Research Consortium 
for the Participating Institutions 
of the SDSS Collaboration including 
the Brazilian Participation Group, 
the Carnegie Institution for Science, 
Carnegie Mellon University, Center for 
Astrophysics | Harvard \& 
Smithsonian, the Chilean Participation 
Group, the French Participation Group, 
Instituto de Astrof\'isica de 
Canarias, The Johns Hopkins 
University, Kavli Institute for the 
Physics and Mathematics of the 
Universe (IPMU) / University of 
Tokyo, the Korean Participation Group, 
Lawrence Berkeley National Laboratory, 
Leibniz Institut f\"ur Astrophysik 
Potsdam (AIP),  Max-Planck-Institut 
f\"ur Astronomie (MPIA Heidelberg), 
Max-Planck-Institut f\"ur 
Astrophysik (MPA Garching), 
Max-Planck-Institut f\"ur 
Extraterrestrische Physik (MPE), 
National Astronomical Observatories of 
China, New Mexico State University, 
New York University, University of 
Notre Dame, Observat\'ario 
Nacional / MCTI, The Ohio State 
University, Pennsylvania State 
University, Shanghai 
Astronomical Observatory, United 
Kingdom Participation Group, 
Universidad Nacional Aut\'onoma 
de M\'exico, University of Arizona, 
University of Colorado Boulder, 
University of Oxford, University of 
Portsmouth, University of Utah, 
University of Virginia, University 
of Washington, University of 
Wisconsin, Vanderbilt University, 
and Yale University. \par

%





\bibliographystyle{aasjournal}
\bibliography{References}



\appendix
\section{The DF modelling}{\label{sec:galaxy}}
\restartappendixnumbering

\begin{figure}[h]
 \includegraphics[width=\textwidth]{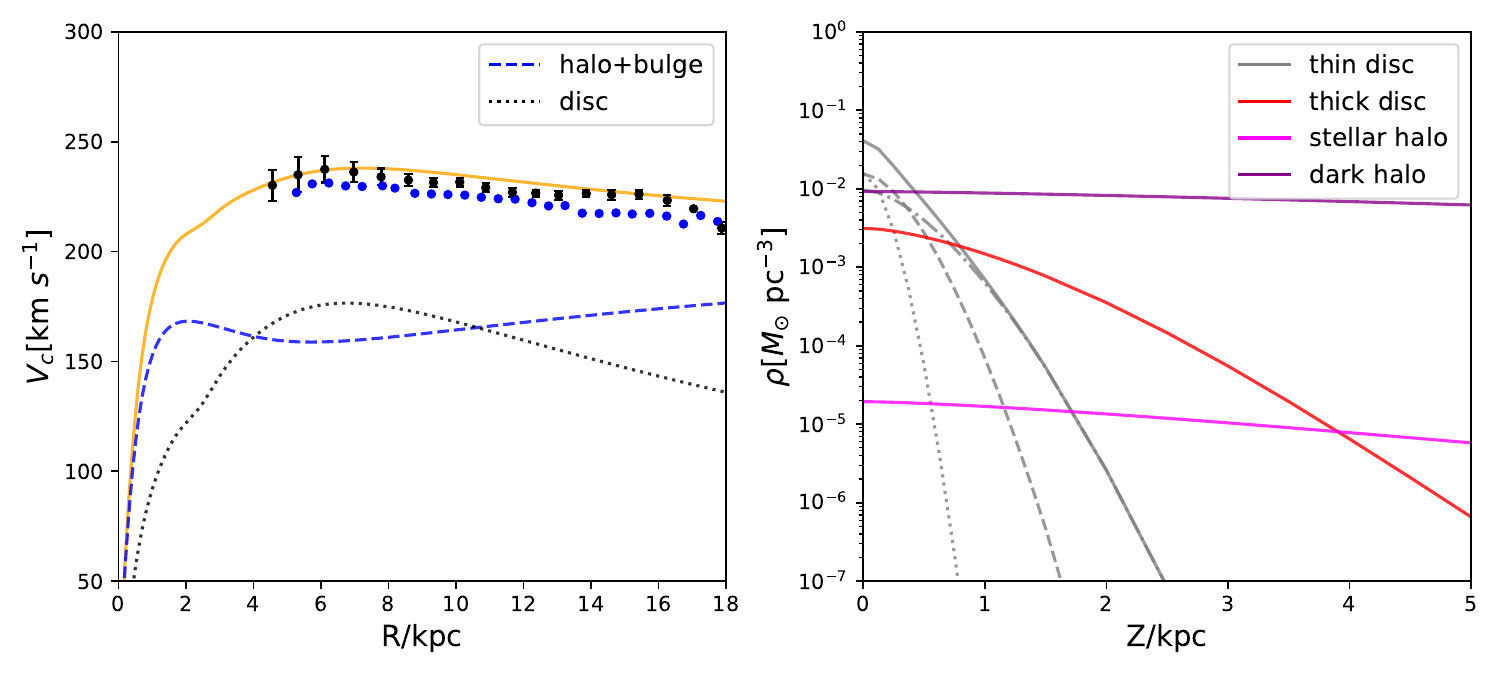}
 \centering
 \caption{The DF modelling framework in this work. The orange curve in the left panel shows the circular velocity predicted by the model, while the blue and black dots represent the measurements of red giants \citep{Eilers2019} and Classical Cepheids \citep{Ablimit2020} respectively. The vertical density of all the stellar components and dark matter halo at solar radius $\rm{R} = 8.2$ are shown in the right panel.}
 \label{fig:circular_velocity}
\end{figure}

In equilibrium, the DF can be taken as a function of the three action variables $(J_R,J_z,J_\phi)$, which serve as isolating integrals of motion \citep{Binney2008}. Following the modeling framework of \citet[hereafter \citetalias{Binney2023}]{Binney2023}, an equilibrium MW model can be defined by a superposition of similar DFs. The DF of each Galactic component is a specified function $f(\vJ)$ of the action integrals. In this model, the disk components are characterized by exponential DFs, which provide a physically reasonable description across action space compared to quasi-isothermal DFs. A double power-law DF is employed to model the bulge, stellar halo, and dark halo. The total gravitational potential of the MW is then derived from the DFs. A detailed description of this axisymmetric Milky Way model can be found in \citet[]{Binney2023,Binney2024}. The circular velocity generated from this axisymmetric potential is shown in the left panel of Figure~\ref{fig:circular_velocity}.

The spheroidal components in the MW, including the bulge and dark/stellar halos, are modeled by a modified double power-law DF, identical to that employed in \citetalias{Binney2023} . 
\begin{equation}
    f(\vJ)=\frac{M}{(2\pi J_{0})^{3}}\frac{(1+[J_{0}/h_{J}]^{\gamma})^{\alpha_{\rm{in}}/\gamma}}
    {(1+[g_{J}/J_{0}]^{\gamma})^{\beta_{\rm{out}}/\gamma}}e^{-(g_{J}/J_{cut})^{\delta}},
    \label{eq:db}
\end{equation}

The modifications of this model compared to \citet{Li2022a} are the linear combinations of $g(\bf{J})$ and $h(\bf{J})$, which endeavors to solve the unphysical behaviour in velocity distributions at small $\left | \rm{v}_\phi \right |$.

This study follows the approach of \citet[hereafter \citetalias{Binney2024}]{Binney2024}, which adopts a truncated exponential DF model for the Galactic disks, expressed as
\begin{equation}
    \label{eq:bulgeDF}
f(\vJ)=f_\phi(J_\phi)f_r(J_R,J_\phi)f_z(J_z,J_\phi)f_{\rm{ext}}(J_\phi)f_{\rm{int}}(J_\phi).
\end{equation}
In this model, the function $f_r$ controls the velocity dispersions $\sigma_R$ and $\sigma_\phi$ near the Galactic plane, while $f_z$ governs both the thickness of the disk and the vertical velocity dispersion $\sigma_z$. The factor $f_\phi$ generates a roughly exponentially declining surface density $\Sigma(R)\simeq\exp(-R/R_\d)$. The functions $f_{\rm ext}$ and $f_{\rm int}$ impose truncations at the outer and inner disk radius, respectively. 
The young thin disk follows a tapered exponential profile with an inner truncation radius, consistent with the model of \citet{Li2022b}. The high-$\alpha$ disk exhibits a sharp density decline at large radii at about $J_\phi \approx 2000 \kms \kpc$. A detailed description of this model can be found in \citetalias{Binney2024}. The vertical density of all the stellar components and dark matter halo at solar radius $R$ = 8.2 are shown in the right panel of Figure~\ref{fig:circular_velocity}. The DF parameters for the disk and spheroidal components are listed in Table~\ref{tab:df} and Table~\ref{tab:df_sph}.

\FloatBarrier
\section{Model parameters}
\restartappendixnumbering

\begin{table*}[h]
    \begin{center}
    \caption{The parameters used for the DF model of the disk components. The units of mass and actions are $10^{10}\msun$ and $\kms\kpc$ respectively. $J_{\phi0}$ determines the radial profile of the disk components. $J_{R0}$ and $J_{z0}$ determine the velocity dispersion profiles in $R$ and $z$ directions. $p_r$ and $p_z$ shapes the sharpness of the velocity dispersion profiles. $D_{\rm ext}$, $J_{\rm ext}$ and $D_{\rm int}$, $J_{\rm int}$ determine the outer and inner truncations respectively. The values of the parameters are based on the fit of APOGEE stars by \citetalias{Binney2024}.}
    \label{tab:df}
    \begin{tabular*}{\textwidth}{@{\extracolsep{\fill}}lcccccccccccc} 
        \hline
        $\rm Name$ &$M$ &$J_{\phi0}$ &$J_{R0}$ &$J_{z0}$ &$J_{\rm ext}$ &$D_{\rm ext}$ &$J_{\rm int}$ &$D_{\rm int}$ &$p_r$ &$p_z$ &$J_{\rm v0}$ &$J_{\rm d0}$ \\ 
        \hline 
        $\rm Young\,disk$  &$0.45$ &$977.9$ &$2.806$ &$1.296$ &$--$ &$--$ &$186.9$ &$278$  &$-0.76$ &$-0.23$ &$152.1$ &$102.1$ \\
        $\rm Mid\,disk$  &$1.2$ &$1030$ &$22.82$ &$3.24$ &$--$ &$--$ &$--$ &$--$ &$-0.23$ &$-0.7$ &$146.4$ &$731.9$ \\
        $\rm Old\,disk$  &$1.47$ &$508$ &$47.14$ &$10.04$ &$--$ &$--$ &$--$ &$--$ &$0.034$ &$-0.043$ &$132.7$ &$150$ \\
        $\rm Thick\,disk$  &$1.3$ &$399$ &$116.4$ &$54.6$ &$2212$ &$207.8$ &$--$ &$--$ &$0.1$ &$0.17$ &$150$ &$10$ \\
        \hline
    \end{tabular*}
    \end{center}
\end{table*}

\begin{table*}[h]
\begin{center}
    \caption{The parameters used for the spheroidal DF model of the bulge and halo. The units of mass and actions are $10^{10}\msun$ and $\kms\kpc$ respectively. $\alpha_{\rm in}$ and $\alpha_{\rm out}$ represent the inner and outer slopes in the double power law model. $J_{\rm cut}$ denotes the outer cut of the halo and bulge and $\delta$ defines the sharpness of the downturn. $J_0$ is the break action which determines the break radius of the inner and outer part. $F_{\rm in}$, $F_{\rm out}$, and $\beta$ determine the shape of the components and the velocity anisotropy.}
    \label{tab:df_sph}
        \begin{tabular*}{\textwidth}{@{\extracolsep{\fill}}lcccccccccc}
            \hline
            $\rm Name$ & $M$ & $\alpha_{\rm in}$ & $\alpha_{\rm out}$ & $J_{\rm core}$ & $J_{\rm cut}$ & $J_{0}$ & $F_{\rm in}$ & $F_{\rm out}$ & $\beta$ & $\delta$\\ 
            \hline 
            $\rm Bulge$ & $1.1$ & $0.5$ & $1.8$	& $5$ & $200$ & $19.5$ & $3$ & $2$ & $0.6$ & $2$\\
            $\rm Halo$ & $0.04$ & $1.6$ & $4.2$ & $6.148$ & $4e+04$	& $583.2$ & $1.8$ & $1.2$ & $0.5$ & $2$\\
            \hline
        \end{tabular*}
\end{center}
\end{table*}

\FloatBarrier
\section{Velocity distribution of the observational sample}
\restartappendixnumbering
\begin{figure*}[h]
 \includegraphics[width=\textwidth]{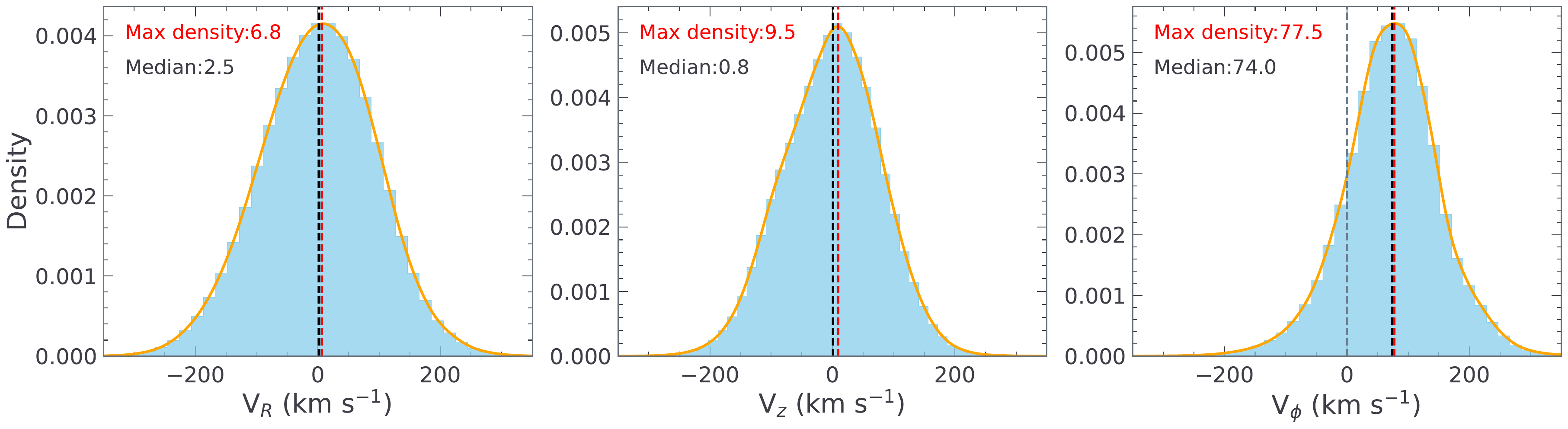}
 \caption{Velocity distribution of the samples with NN prediction values in the selected region. Kernel density plots are superimposed on the histograms. The grey dashed line represents zero velocity. The velocity at the point of maximum kernel density and the median velocity are indicated by the red and black dashed lines, respectively, and are displayed in the upper left corner of each subplot.}
 \label{fig:velocity_obs}
\end{figure*}
\end{CJK*}
\end{document}